\documentclass[a4paper,11pt]{article}
\pdfoutput=1 

\usepackage{jinstpub} 

\title{Characterization of Hamamatsu 14160 series of Silicon Photo-Multipliers}

\author[a,1]{P.W.~Cattaneo\note{Corresponding author.},} 
\author[a,b]{A.~Menegolli,}
\author[a]{M.C.~Prata,}
\author[a]{G.L.~Raselli,}
\author[a]{M.~Rossella,}

\affiliation[a]{INFN Sezione di Pavia, Pavia, Italy} 
\affiliation[b]{Dipartimento di Fisica, University of Pavia, Pavia, Italy} 

\emailAdd{paolo.cattaneo@pv.infn.it}

\abstract{Silicon Photo-Multipliers (SiPMs) are semiconductor-based photo-detectors with performances similar to the traditional Photo-Multiplier Tubes (PMTs). An increasing number of experiments dedicated to particle detection in colliders, accelerators, astrophysics, neutrino and rare-event physics involving scintillators are using SiPMs as photodetectors. They are gradually substituting PMTs in many applications,  especially where low voltages are required and high magnetic field is present. Hamamatsu Photonics K.K., one of leading producers of  photo-detectors, in the last year introduced the S14160 series of SiPMs with improved performances. In this work, a  characterization of these devices will be presented in terms of breakdown voltages, pulse shape, dark current and gain. Particular attention has been dedicated to the analysis of the parameters as function of temperature.}

\keywords{}

\begin{document}
\maketitle
\flushbottom

\section{Introduction}
In latest years, solid state photo-detectors have become a staple for many experiments in high energy physics, either at the accelerators or
devoted to neutrino and rare-event physics. In particular, Silicon Photo-Multipliers (SiPMs) are semiconductor-based photo-detectors adopting the avalanche photo-diode approach to get performance similar to the traditional Photo-Multiplier Tubes (PMTs). A number of neutrino and Dark Matter experiments \cite{dune} \cite{darkside} plan to use SiPM also at cryogenic temperatures as scintillation light detectors. Despite the smaller window surface with respect to traditional large area PMTs, SiPMs operating at cryogenic temperature have the advantage of a thermal component of the noise decreasing with the reduction of the temperature, thus allowing the use of these devices at higher overvoltage \cite{bib.sipm1} \cite{bib.sipm2}. Hamamatsu Photonics K.K., one of leading producers of photo-detectors, in the last year introduced the S14160 series of SiPMs with improved performances. The aim of this work is the test these new devices in terms of breakdown voltages, pulse shape, dark current and gain, and compare them with previous series. 
An experimental apparatus based on a climatic chamber allowed to carry out the measurements in a temperature range from -40$^\circ$C to +40$^\circ$C.

\section{Experimental apparatus}
\label{sec:apparatus}

\begin{figure}[htbp]
\centering
\includegraphics[width=.6\textwidth]{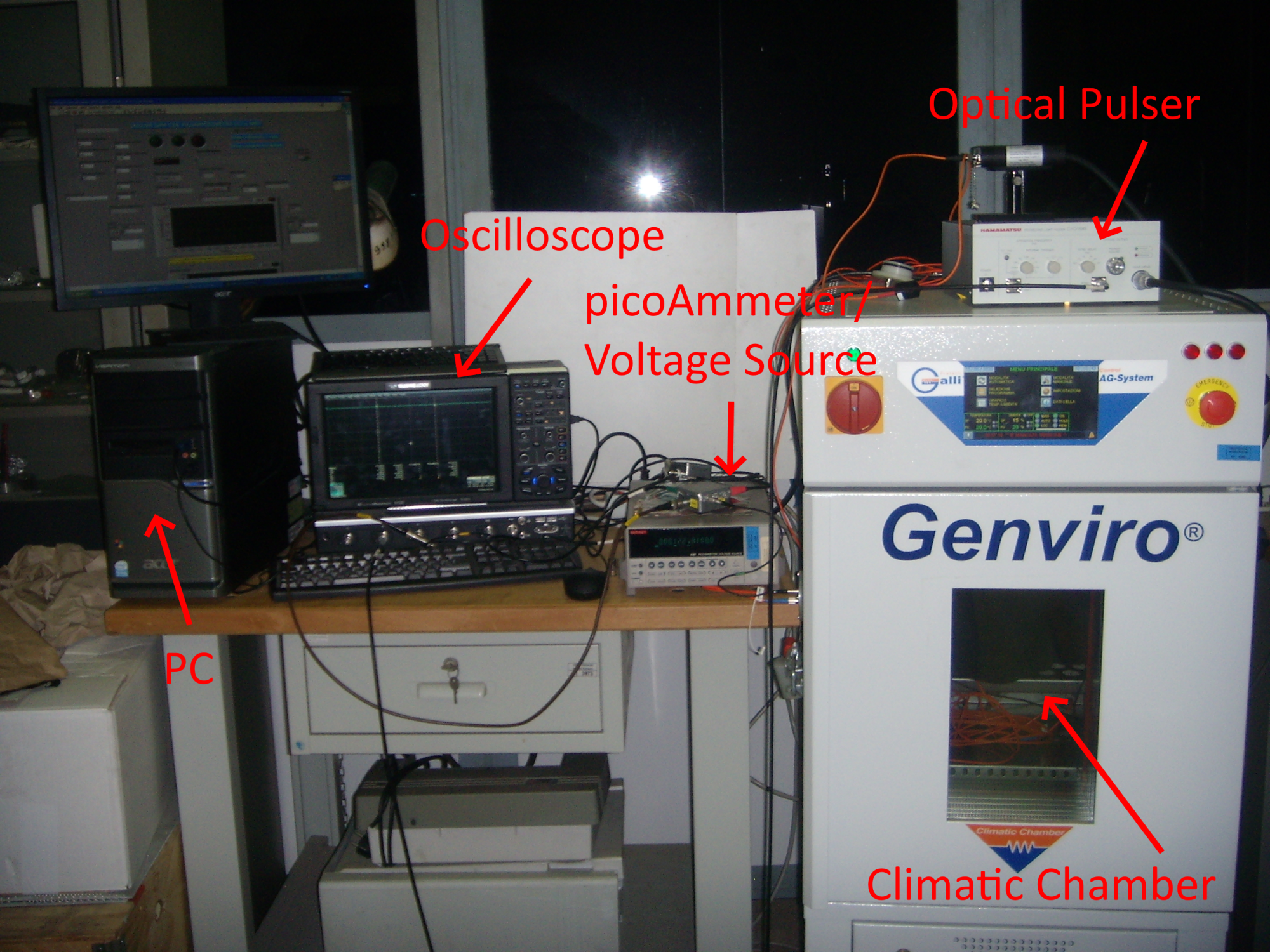}
\caption{Experimental set-up used for the characterization of the S14160 SiPM series.}
\label{fig:apparatus}
\end{figure}

\begin{figure}[htbp]
\centering
\includegraphics[width=.6\textwidth]{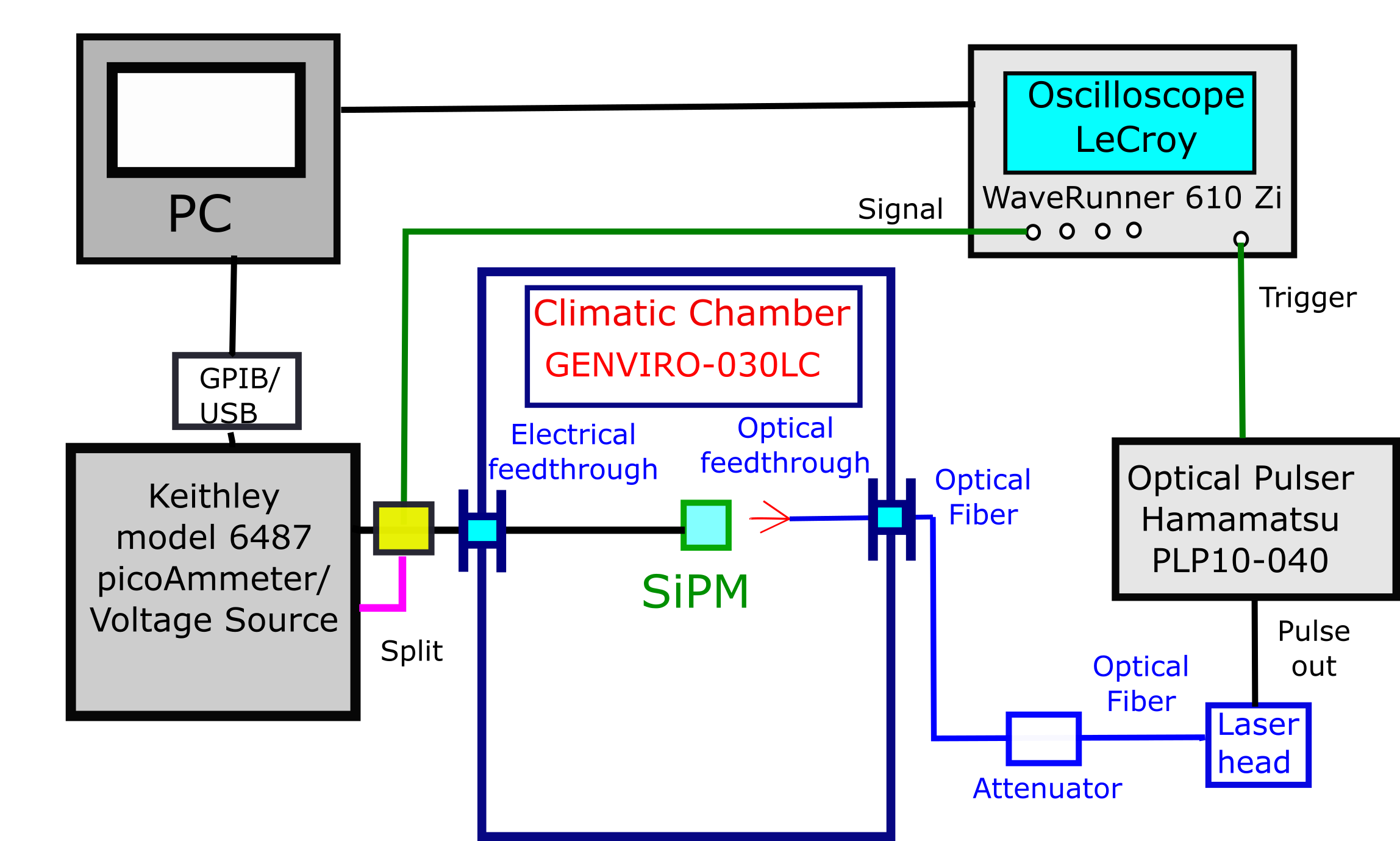}
\caption{Scheme of the experimental set-up.}
\label{scheme}
\end{figure}

\begin{figure}[htbp]
\centering
\includegraphics[width=.45\textwidth]{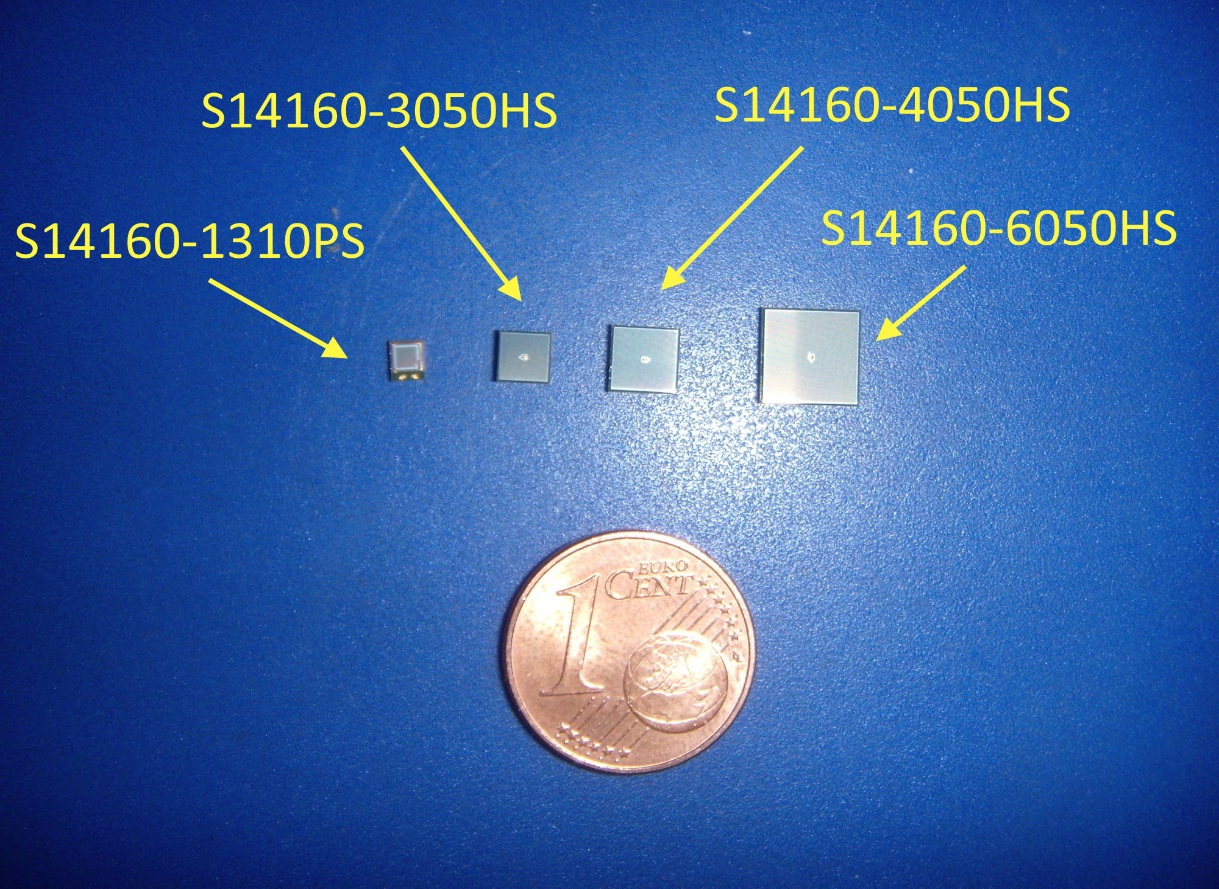}
\includegraphics[width=.53\textwidth]{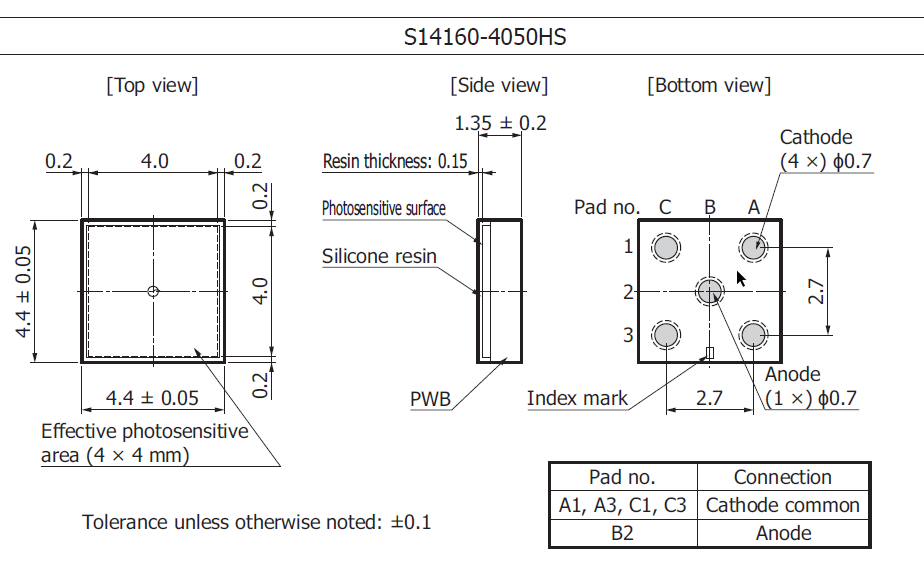}
\caption{Picture of the devices under test.}
\label{sipm}
\end{figure}

The adopted experimental setup is based on a climatic chamber (F.lli Galli model Genviro-030LC) with a temperature range from $-60^\circ$C to  $+60^\circ$C that can house the SiPM to be tested. The inner part of the chamber can be connected to the external world by means of  electrical and optical (optical fiber) feed-throughs.  
Fig.~\ref{fig:apparatus} shows a picture of the chamber.
A Keithly 6487 model picoAmmeter/Source-generator is used both to apply the voltage to the SiPM and to measure the current.
An Hamamatsu optical pulser, model PLP10-040, with an emission wavelength $\lambda$ = 405 nm, pulse width of 60 ps (FWHM) and peak power of 200 mW, is used to illuminate the device. About $10^7$ photons per pulse are directly injected in a FC optical fiber followed by an optical fiber attenuator that can reduce the number of photons
from $10^7$ to zero. The optical pulser can pulse up to 100MHz, but all the measurements
of this work have been carried out at 10Hz.
Signals are acquired by means of a LeCroy digital Oscilloscope model Waverunner 610Zi, with 8-bit vertical resolution, 1~GHz bandwidth, and 20~GSa/s sampling rate. The complete set-up is shown in Fig.~\ref{scheme}. Several Hamamatsu devices from 14160 series have been selected to be tested (see Fig.~\ref{sipm}): 

\begin{itemize}
\setlength\itemsep{0pt}
    \item S14160-3050HS: 3$\times$3 mm$^2$ active area, 50 $\mu$m cell, 3531 pixels, 74\% fill factor, silicon window.
    \item S14160-4050HS: 4$\times$4 mm$^2$ active area, 50 $\mu$m cell, 6331 pixels, 74\% fill factor, silicon window.
    \item S14160-6050HS: 6$\times$6 mm$^2$ active area, 50 $\mu$m cell, 14331 pixels, 74\% fill factor, silicon window.
    \item S14160-1310PS: 1.3$\times$1.3 mm$^2$ active area, 10 $\mu$m cell, 16675 pixels, 31\% fill factor, silicon window.
    \item S14160-1315PS: 1.3$\times$1.3 mm$^2$ active area, 15 $\mu$m cell, 7296 pixels, 49\% fill factor, silicon window.
\end{itemize}

Hamamatsu S14160 series are innovative surface mount SiPMs with an higher PDE, a lower temperature coefficient and a lower operation (breakdown) voltage in comparison to previous Hamamatsu devices. They are immune to effects of magnetic fields, are poorly affected by cross-talk and after-pulses, operate at a typical breakdown voltage of 38V and have an excellent time resolution. The 3$\times$3 mm$^2$, 4$\times$4 mm$^2$ and 6$\times$6 mm$^2$ version have 50$\mu$m cells, a PDE up to 50\% (at the peak wavelength and at 2.7V overvoltage) and a typical gain of $10^6$. They are suited for scintillation detectors and for industrial applications such as for PET, radiation monitor etc.. The higher PDE is due to the HWB (Hole Wire bonding) technology allowing for small dead space in the photosensitive area. The 1.3$\times$1.3 mm$^2$ versions have smaller pixel size (10$\mu$m and 15$\mu$m), high fill factor and wide dynamic range; they show low cross-talk and after-pulses and a typical gain of $10^5$.
They are suggested for high energy physics experiments, fluorescence measurement,
flow cytometry, DNA sequencers and environmental analysis.

\section{Results}
\begin{figure}[htbp]
\centering
\includegraphics[width=.61\textwidth]{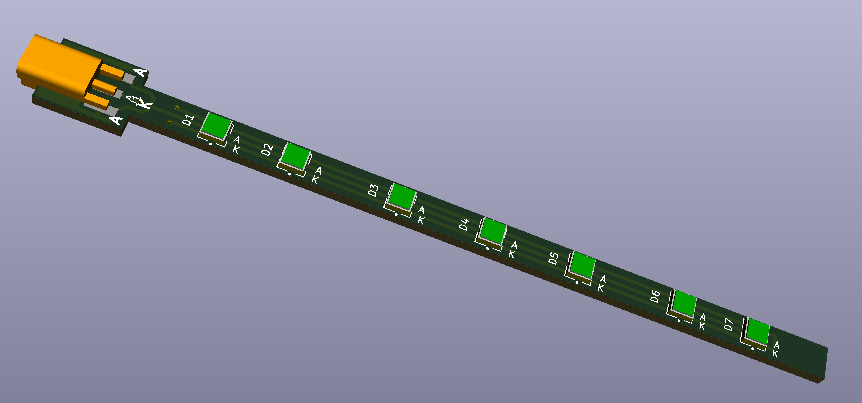}
\includegraphics[width=.38\textwidth]{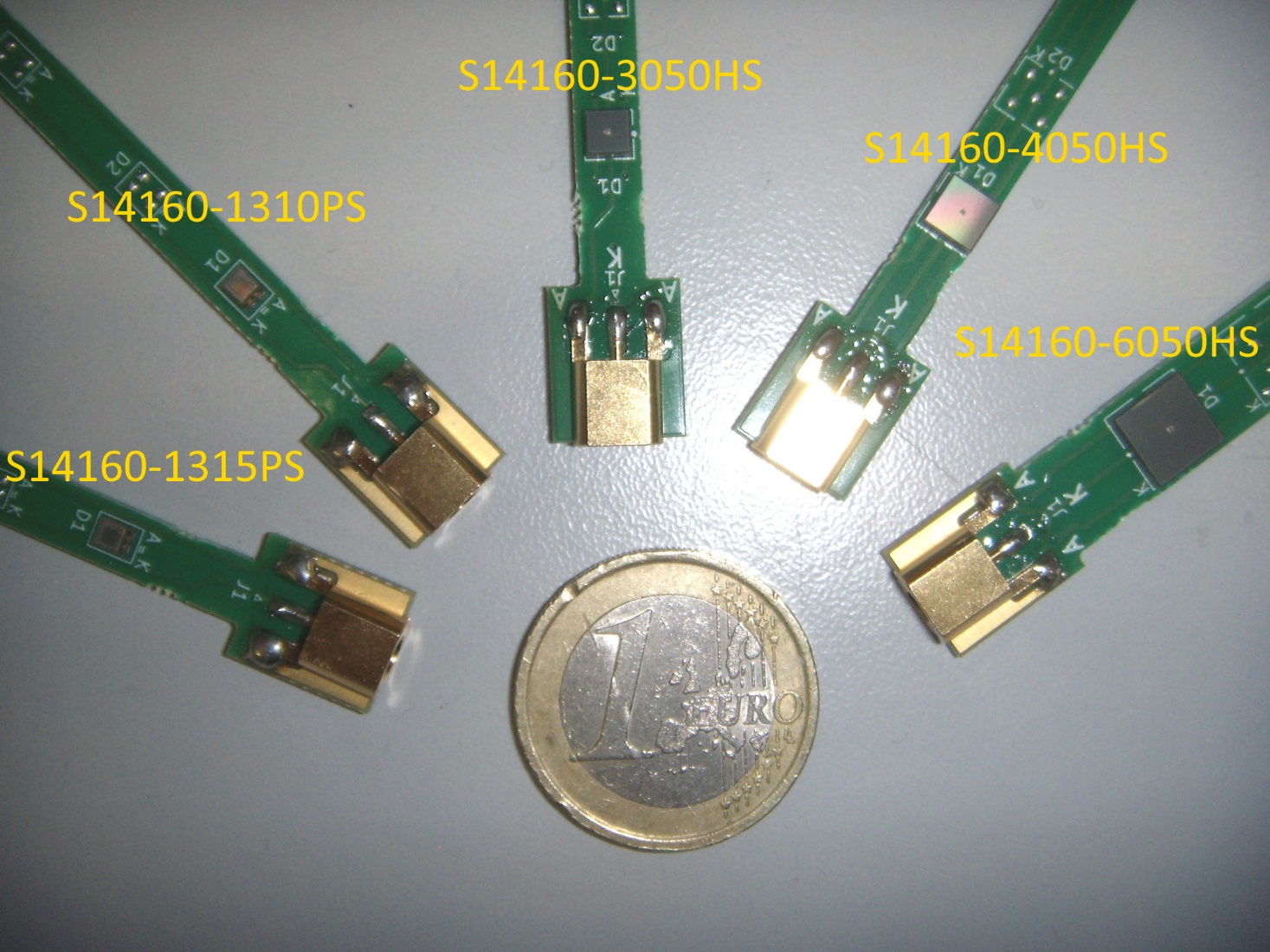}
\caption{Left: 3D rendering of our custom board for SiPM mounting. Right: the  electronic boards housing the SiPM models under test.}
\label{board}
\end{figure}

Considering that all the devices under test are surface-mount technology,
we decided to solder them on electronic boards each equipped with a coaxial MCX connector, as shown in Fig.~\ref{board}.
Soldering was performed after a vacuum baking of 48 hours at 60$^\circ$C, by means of Vapor Phase Reflow Soldering process that guarantees a uniform and well controlled temperature of the process and consequently no damage on the plastic window of the devices.

\begin{figure}[htbp]
\centering
\includegraphics[width=.45\textwidth]{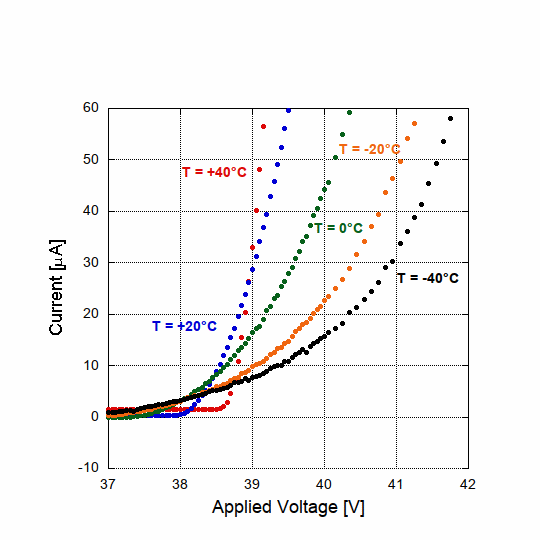}
\includegraphics[width=.45\textwidth]{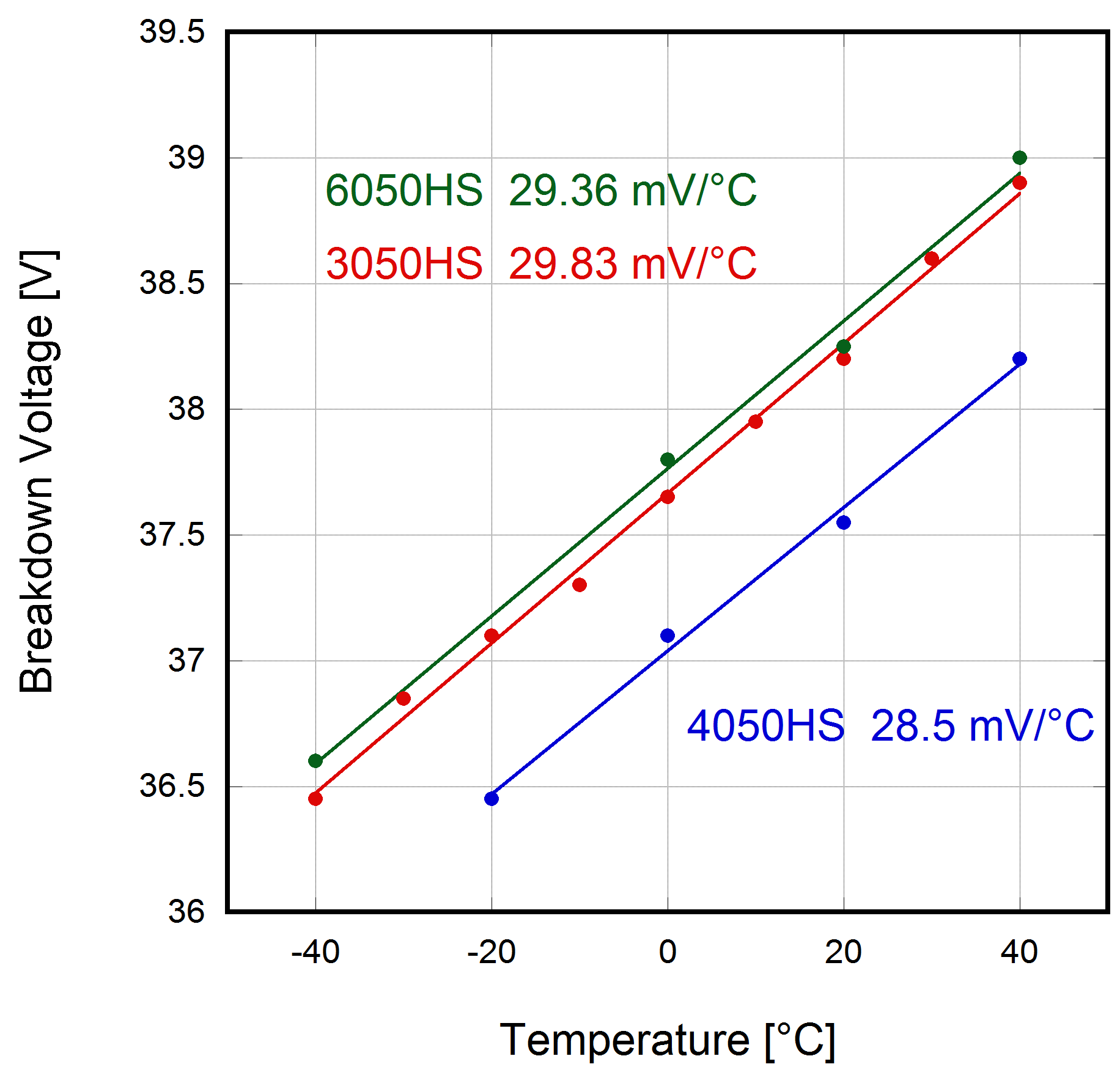}
\caption{Left: I-V curve at different temperatures for  S14160-3050HS. Right: Breakdown voltage as a function of temperature.}
\label{ivbd}
\end{figure}

\begin{figure}[htbp]
\centering
\includegraphics[width=.49\textwidth]{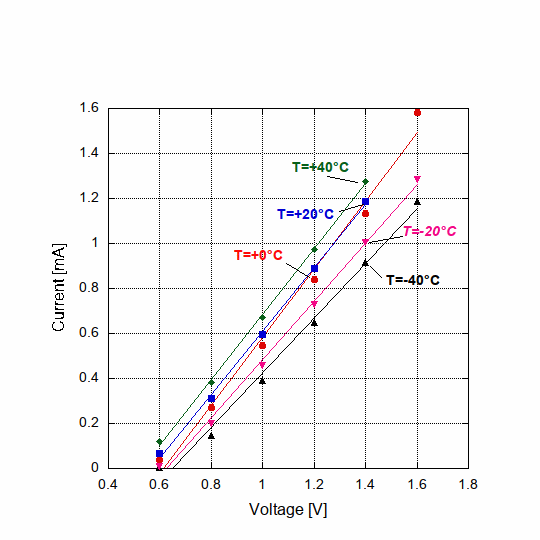}
\includegraphics[width=.49\textwidth]{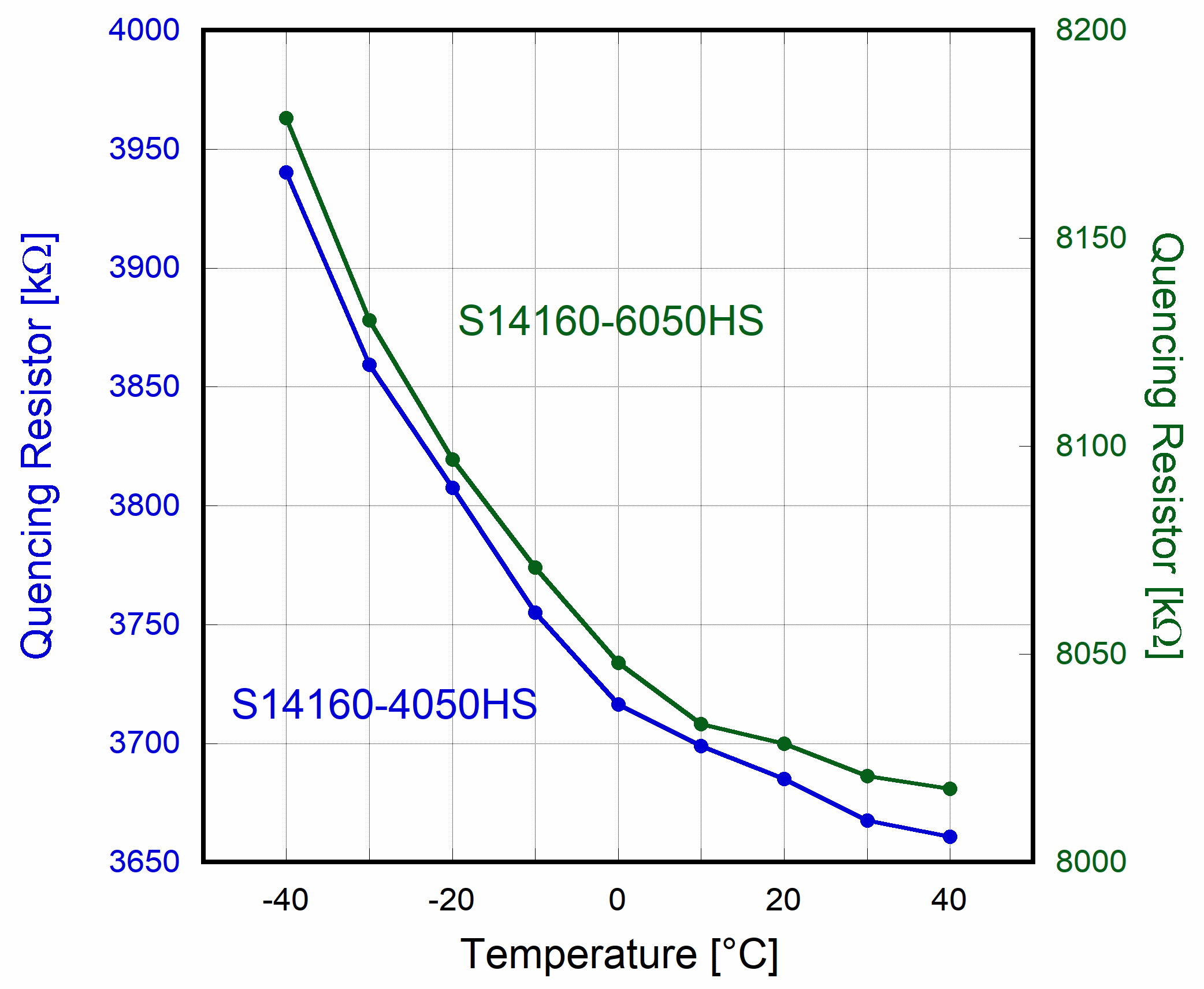}
\caption{Left: Direct I-V curve for S14160-3050HS at several temperatures. Right: quenching  resistor as function of temperature for S14160-4050HS and S14160-6050HS.}
\label{quench}
\end{figure}

Preliminary results have been obtained from tests on the devices described above. Some of them are here reported in terms of:

\begin{itemize}
\setlength\itemsep{0pt}
    \item  I-V curve of S14160-3050HS for various temperatures plotting dark current versus bias voltage (Fig.~\ref{ivbd} left); the breakdown voltage is defined as the voltage where the second derivative of the curves peaks, 
    \item breakdown voltage as function of temperature with temperature coefficient displayed for the three 50 $\mu$m cell SiPMs
    (Fig.~\ref{ivbd} right): all models 
    show a linear behaviour, with similar temperature coefficients $\sim$30 mV/$^\circ$C. The same measurement
    performed in the past with previous S12572 series SiPMs (3$\times$3 mm$^2$ area, 50 $\mu$m cell) gave a temperature
    coefficient of $\sim$ 60 mV/$^\circ$C \cite{tc12572}, 
    thus showing the larger stability of the new series;
    \item Direct I-V curve for S14160-3050HS at several temperatures (Fig.~\ref{quench} left);
    \item quenching resistor for S14160-6050HS and 4050HS as function of temperature (Fig.~\ref{quench} right).
\end{itemize}

\begin{figure}[htbp]
\centering
\includegraphics[width=.45\textwidth]{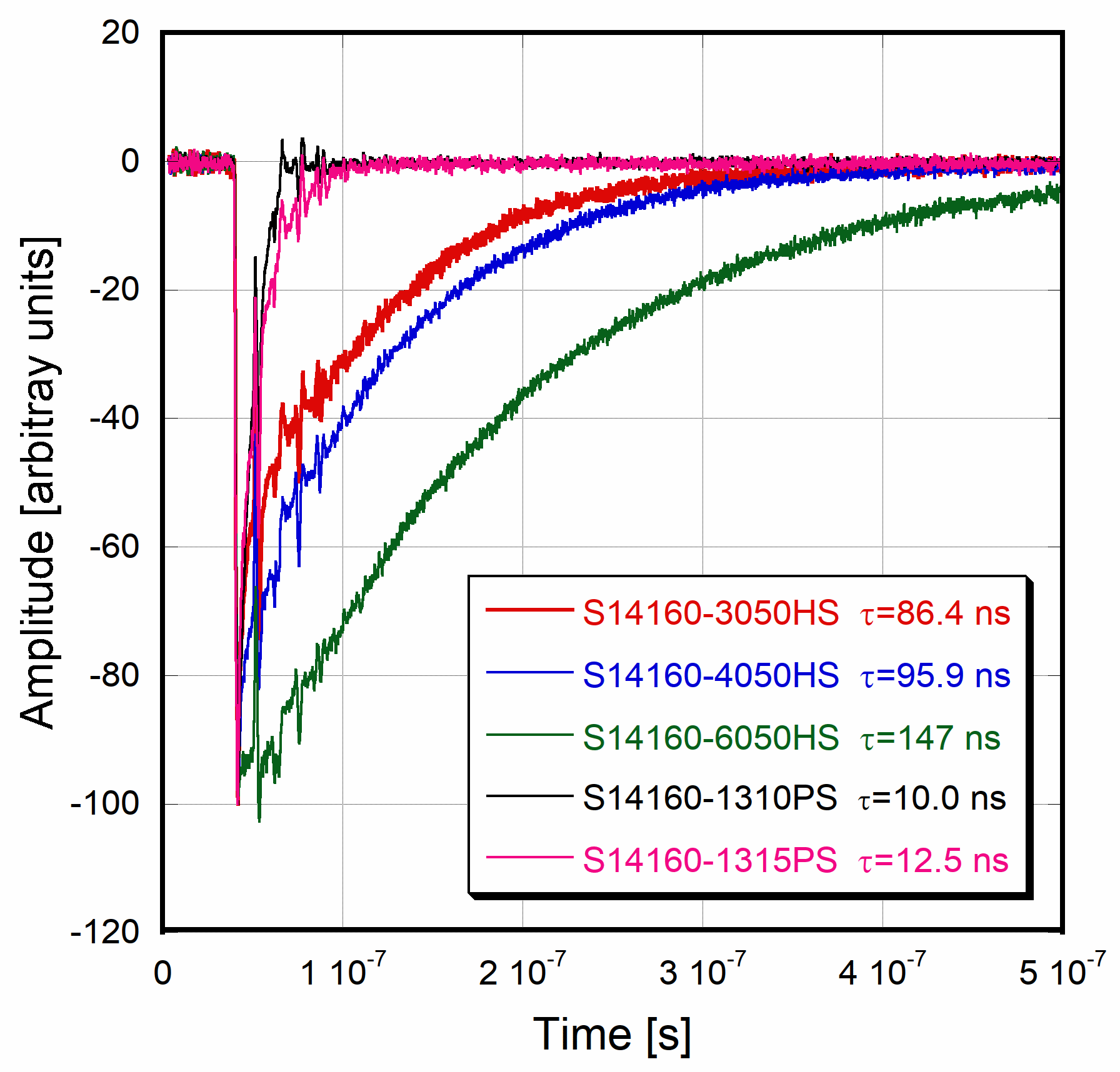}
\includegraphics[width=.45\textwidth]{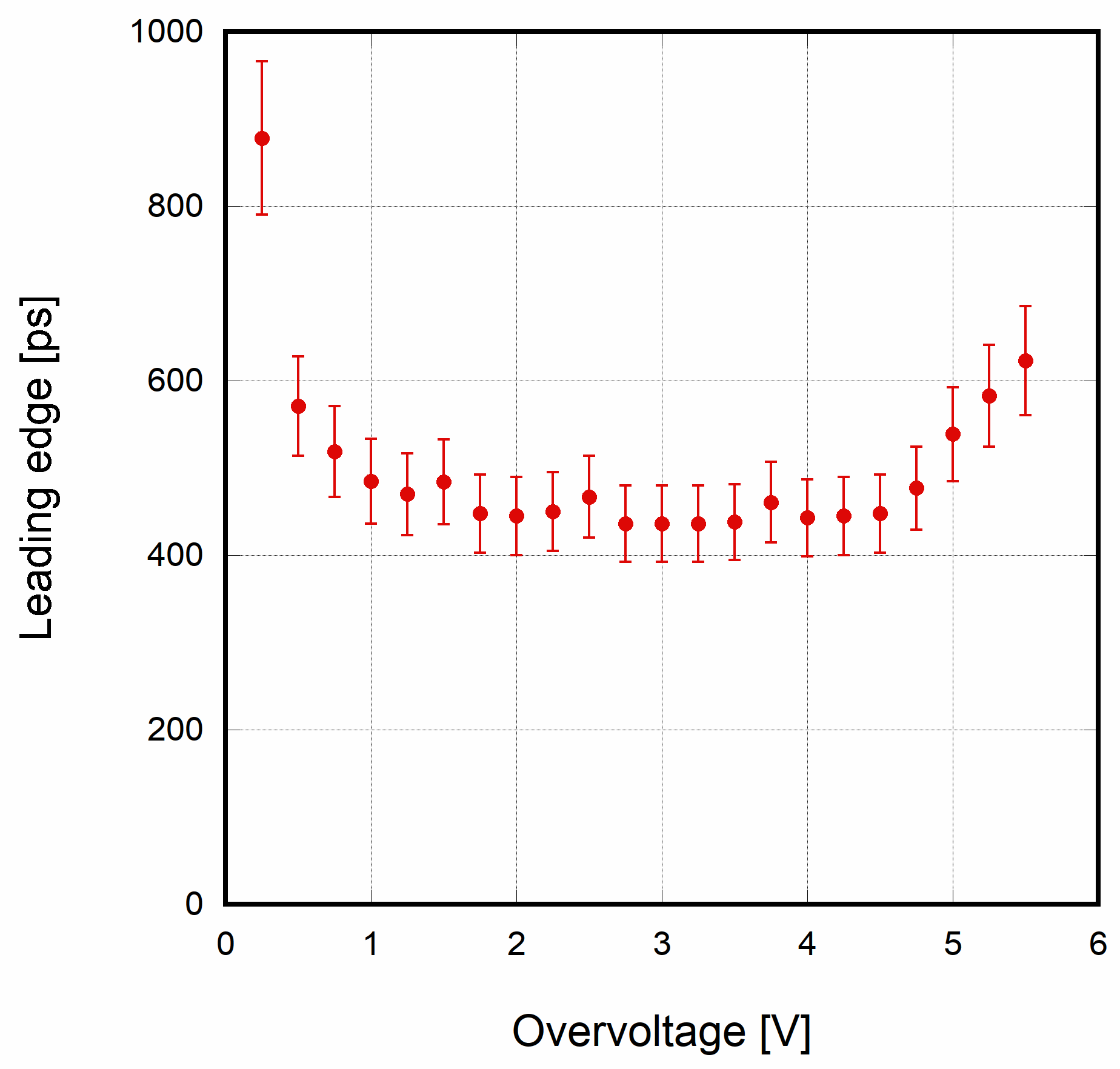}
\caption{Left: typical pulse shape for all devices under test with the indication of  the decay time. Right: leading edge of S14160-3050HS as a function of the overvoltage.}
\label{leps}
\end{figure}

A further comparison with the S12572 series has been carried out first of all in terms of leading edge and decay time of SiPM signals 
under illumination provided by the Hamamatsu optical pulser (see Fig.~\ref{scheme}):
the typical pulse shapes of all devices under test are shown in Fig.~\ref{leps}-left, while in Fig.~\ref{leps}-right the rise time of the S14160-3050HS as function of the overvoltage is shown.

\begin{figure}[htbp]
\centering
\includegraphics[width=.48\textwidth]{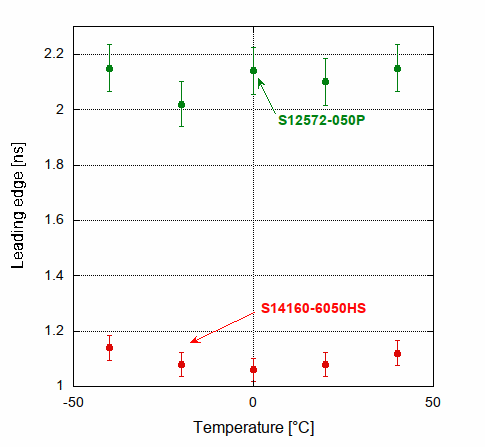}
\includegraphics[width=.49\textwidth]{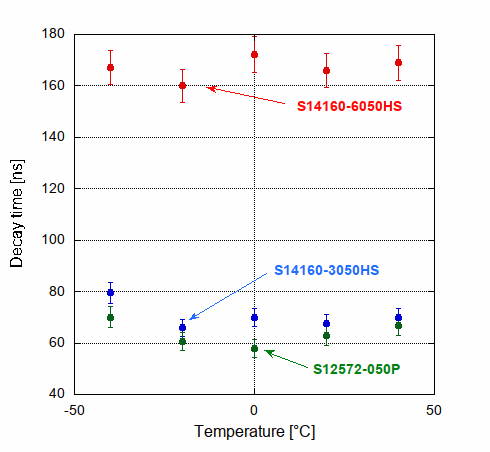}
\caption{Comparison between S14160 and S12572 leading edge (left) and decay time (right) as a function of the temperature.}
\label{ledt}
\end{figure}

Fig.~\ref{ledt} shows the results on the leading edge (left) and on the decay time (right) of the SiPM signal for the  S14160 and S12572 series. The leading edge of the S14160-6050HS SiPM is about the half of the one of the S12572 SiPM series. It can be noted that for both series the leading edge is poorly affected by the temperature change, being stable in the range from -40$^\circ$C to +40$^\circ$C defined within the climatic chamber. The decay time is instead similar for the S14160 and S21572 series when the SiPMs have the same 3$\times$3 mm$^2$ area. The S14160-6050HS, with a 6$\times$6 mm$^2$ area, is affected by a longer decay time, as expected due to the higher quenching resistor.

\begin{figure}[htbp]
\centering
\includegraphics[width=.48\textwidth]{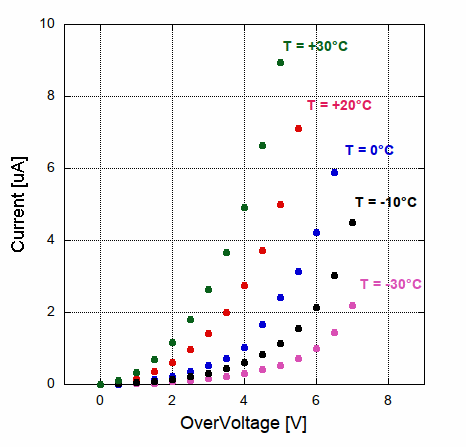}
\includegraphics[width=.49\textwidth]{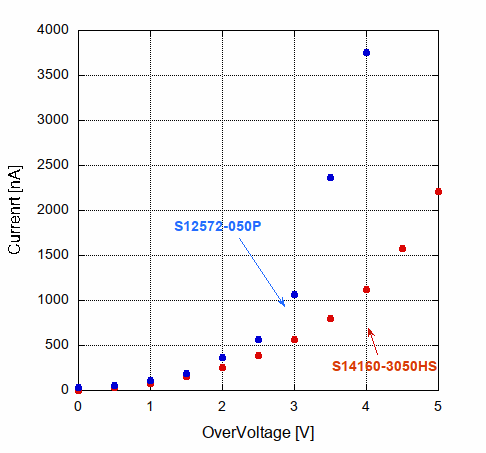}
\caption{Left: dark current of S14160-6050HS as a function of the overvoltage, for several temperatures. Right: comparison between S14160-3050HS and S12572-050P in terms of Dark Current as a function of the overvoltage.}
\label{dcov}
\end{figure}

In order to characterize the devices in terms of noise, we measured the dark current of S14160-6050HS as a function of the overvoltage for several temperatures, as shown in Fig.~\ref{dcov}-left. Comparing the new series with the S12572, a large reduction of the noise for a same overvoltage value is seen, as shown in Fig.~\ref{dcov}-right for a temperature of 30$^\circ$C. Both SiPMs are 3$\times$3 mm$^2$ area, 50 $\mu$m cell.

\begin{figure}[htbp]
\centering
\includegraphics[width=.49\textwidth]{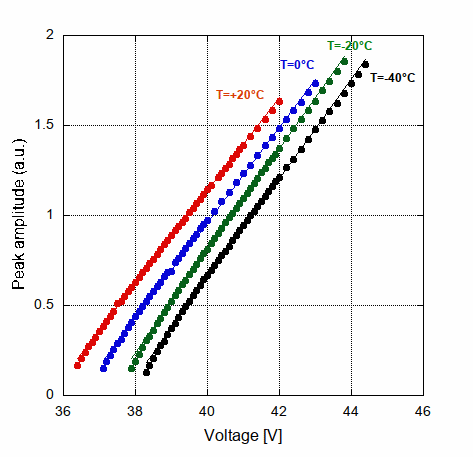}
\includegraphics[width=.48\textwidth]{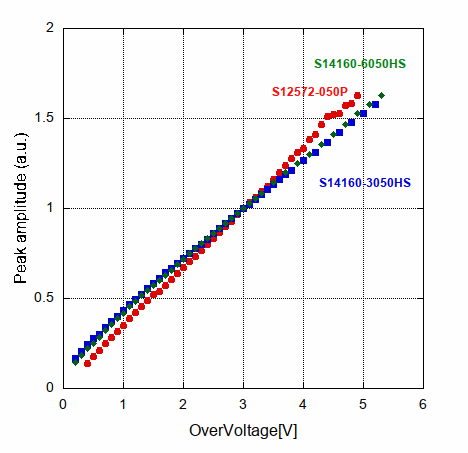}
\caption{Left: peak amplitude of S14160-3050HS as a function of the applied voltage, measured at several temperature. Right: peak amplitude of S14160 and S12572 series as a function of the overvoltage, measured at room temperature.}
\label{peak}
\end{figure}

Finally, a preliminary study was performed on the signal amplitude of several SiPMs, illuminated by the Hamamatsu optical pulser with a fixed number of photons per pulse, by evaluating the peak directly from the oscilloscope output and varying the overvoltage. Results are shown in Fig.~\ref{peak} in terms of pulse peak normalized to the value measured for an overvoltage of 3 Volts. In the figure, the quantity displayed in the vertical axis is  proportional to the product PDE $\times$ Gain. 
If we consider a constant PDE as a function of the overvoltage, the curve slope corresponds to the relative variation $\frac{\Delta G}{\Delta V}$ of the gain with respect to the supplied voltage. S14160-6050HS SiPM has an almost linear behaviour from -40$^\circ$C to room temperature (Fig.~\ref{peak}-left). The relative variation of the peak amplitude is the same for all the temperatures, being the slopes of all curves identical. In Fig.~\ref{peak}-right the S14160 and S12572 series are compared at room temperature: while the two S14160 models have the same relative variation, the slope of the S12572 curve (red dots) is slightly larger. 

\section{Conclusions}

The Hamamatsu S14160 series Silicon PM has been widely tested in terms of I-V curve, voltage breakdown temperature coefficient, quenching resistor and pulse response as a function of temperature, from -40$^\circ$C to +40$^\circ$C.
The performance on the most significant parameters of the S14160 series of SiPM is largely improved with respect to preceding series, in particular in terms of a lower temperature coefficient, a lower breakdown voltage, better noise and higher PDE and gain.

\end{document}